\documentclass[%
 reprint,
superscriptaddress,
nobibnotes,
aps,
prb,
]{revtex4-2}
\usepackage{amsfonts}
\usepackage{amsmath}
\usepackage{amssymb, bm, mathrsfs}
\usepackage{graphicx, xcolor}
\usepackage{soul}
\usepackage{natbib} 
\usepackage{url}
\usepackage[labelformat=simple]{subcaption}
\usepackage{physics}
\usepackage{chemformula}
\usepackage[colorlinks=true,citecolor=blue]{hyperref}
\usepackage[capitalise]{cleveref}
\usepackage{appendix, comment}
\usepackage[mathlines]{lineno}
\usepackage{ragged2e}
 
\DeclareCaptionJustification{justified}{\justifying}
\begin{document}

\title{Thermal-Field Electron Emission from Three-Dimensional \\Topological Semimetals}

\author{Wei Jie Chan}
\affiliation{Science, Mathematics and Technology (SMT), Singapore University of Technology and Design (SUTD), 8 Somapah Road, Singapore 487372}

\author{Yee Sin Ang}
\email{yeesin\_ang@sutd.edu.sg}
\affiliation{Science, Mathematics and Technology (SMT), Singapore University of Technology and Design (SUTD), 8 Somapah Road, Singapore 487372}

\author{L. K. Ang}
\email{ricky\_ang@sutd.edu.sg}
\affiliation{Science, Mathematics and Technology (SMT), Singapore University of Technology and Design (SUTD), 8 Somapah Road, Singapore 487372}

\begin{abstract}
A model is constructed to describe the thermal-field emission of electrons from a three-dimensional ($3$D) topological semimetal hosting Dirac/Weyl node(s). 
The traditional thermal-field electron emission model is generalised to accommodate the $3$D non-parabolic energy band structures in the topological Dirac/Weyl semimetals, such as cadmium arsenide (\ch{Cd3As2}), sodium bismuthide (\ch{Na3Bi}), tantalum arsenide (\ch{TaAs}) and tantalum phosphide (\ch{TaP}). 
Due to the unique Dirac cone band structure, an unusual dual-peak feature is observed in the total energy distribution (TED) spectrum.
This non-trivial dual-peak feature, absent from traditional materials, plays a critical role in manipulating the TED spectrum and the magnitude of the emission current.
At zero temperature limit, a new scaling law for pure field emission is derived and it is different from the well-known Fowler-Nordheim (FN) law.
This model expands the recent understandings of electron emission studied for the Dirac $2$D materials into the $3$D regime, and thus offers a theoretical foundation for the exploration in using topological semimetals as novel electrodes. 
\end{abstract}

\maketitle

\section{Introduction \label{sec:Intro}}
Topological Weyl/Dirac semimetals (WSM/DSM)s, a subset of Dirac materials have been studied rapidly over the past decade \cite{Son2006,Oka2009,Xu2011,Lv2015,Mcc2017,Liu2014,Liu2014a,Crassee2018,Wehling2014} due to its electronic \cite{McCann2013,Wan2011,Xiong2016,Wang2017}, optical \cite{Lasia2014,Xu2016,Polatkan2019,Wang2017} and magnetic \cite{Choi2011,Polatkan2019,Wang2017} properties. 
The unconventional band structures about its Dirac point have bought about many interesting applications in electronics \cite{Li2016,Zhu2017,Wang2017a}, spintronics \cite{Smejkal2018}, photonics \cite{Khanikaev2013,Yu2019}, nonlinear optics \cite{Lim2020,Zhang2019,Lim2020a} and topological electronics (topotronics) \cite{Wang2020}. 
Apart from these applications, the physics of electron emission from Dirac materials like carbon-based nanomaterials \cite{Hofmann2003,Liang2008,Zhou2019} or graphene \cite{Giubileo2018,Chen2017,SunAPL2011,AngMRS2017,Wei2012,LiangPRAp2015,AngPRL2018,AngPRAp2019,Huang2017,Ang2021} have also received attentions over the past two decades. 

Field-induced electron emission describes the quantum tunnelling of electrons from a material surface into a vacuum under a strong electric field.
The most well-known field emission physics is described by the Fowler-Nordheim (FN) based models developed for traditional bulk materials \cite{Forbes2006,Forbes2007,Jensen2006}.
Compared to the conventional field emitters \cite{Swason1966,Dionne2009,Chung1994,Binh1996,Young1959,Gadzuk1973}, novel quantum materials exhibit Dirac conic band structure about its Fermi level  with non-parabolic energy dispersion \cite{Wehling2014}, and they have also been experimentally shown to exhibit large field enhancement and stable current emission \cite{Lee2012,DeJonge2005,Li2012}. 
However, the physics of thermal-field emission of some recently discovered topological solids (such as $3$D Dirac/Weyl semimetals) has not been studied in details, which immediately leads to the following questions:
(i) How can the conventional thermal-field emission model be generalized to accommodate the nonparabolic band structure of topological semimetals? 
(ii) How does the Dirac conic band structure affect the themral-field emission behaviours of $3$D WSM/DSM? 
(iii) What are the differences between conventional metal and WSM/DSM in terms of their current, voltage and temperature scaling laws at the both field and thermal-field emission regimes? 

\begin{figure*}
        \begin{subfigure}[t]{1\textwidth}
        \includegraphics[width=1\textwidth]{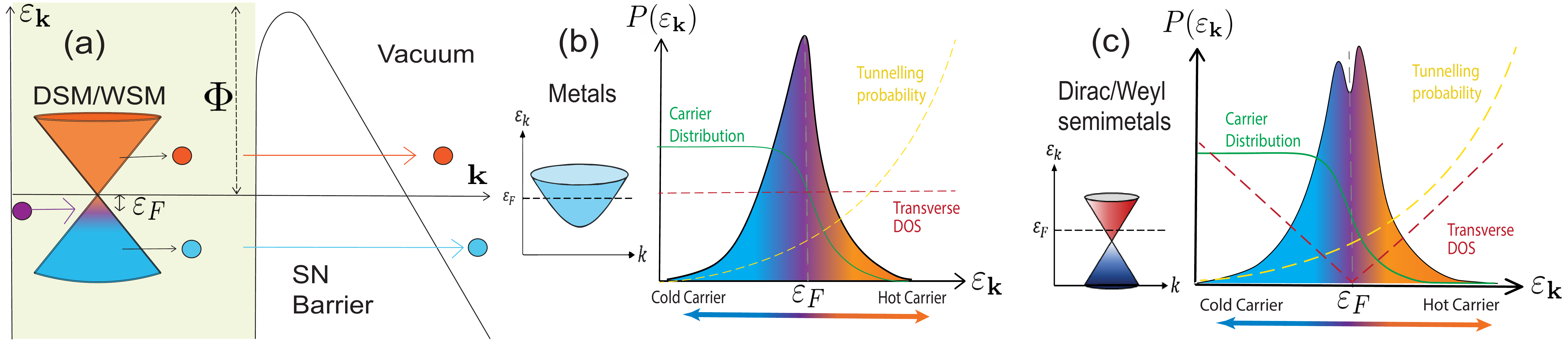}
        \phantomcaption
        \label{fig:1a}
    \end{subfigure}
    \begin{subfigure}[t]{0\textwidth}
        \includegraphics[width=0\textwidth]{Fig1.pdf}
        \phantomcaption
        \label{fig:1b}
    \end{subfigure}
    \begin{subfigure}[t]{0\textwidth}
        \includegraphics[width=0\textwidth]{Fig1.pdf}
        \phantomcaption
        \label{fig:1c}
    \end{subfigure}
    \caption{(a) Schematic diagram of a material with a linear dispersion under the Dirac cone approximation. Under an electric field, the electrons emit through a Schottky-Nordheim (SN) barrier \cite{Murphy1956} with work function, $\Phi$, into vacuum. Orange/blue electrons are emitted respectively, from the conduction/valence bands. The emitted electrons are then replenished at the Fermi level, $\varepsilon_F$ immediately by replacement electrons (purple) entering the Dirac cone \cite{Chung1994}. The relationship of the energy dispersion, carrier density (green solid line), tunnelling probability (yellow dotted line), and in particular, the transverse density of states (DOS) (red dotted line), help sculpt the total energy distribution (TED), $P(\varepsilon_\mathbf{k})$ of (b) a conventional metal with a parabolic dispersion, giving us a single peak TED, and (c) of a DSM/WSM using the Dirac cone approximation, giving us a dual-peak feature in the TED.\label{fig:1}}
\end{figure*}

In this paper, we address the above questions by constructing a generalized thermal-field emission model for the newly discovered $3$D DSMs/WSMs. 
In particular, the total energy distribution (TED) and the emission current density are calculated. 
Our model applies the Dirac cone approximation for DSMs/WSMs and considers the Schottky-Nordheim (SN) barrier \cite{Forbes2006,Murphy1956} at the material-vacuum interface, as seen in \cref{fig:1a}.
The electron emission can reside from either the conduction (orange) or valence (blue) bands and are then replenished (purple) at the intrinsic Fermi level, $\varepsilon_{F0}$. 
The differing (linear) energy dispersion from a Dirac cone allows us to study the thermal-field emission of $3$D DSMs/WSMs.
For a linear dispersion, we predict an unconventional TED behaviour and a new $F^3$ scaling law at the zero temperature limit, which is different from the FN law.
By tuning the applied electric field strength, temperature and Fermi level of a DSM/WSM, we can manipulate the energy profile of TED and the magnitude of the emission current density.
These findings can pave the way for the theoretical study of other topological materials, in particular, materials that are described by Dirac cone(s) and Weyl nodes in the band structure, which cannot be modelled by the traditional thermal-field emission models \cite{Murphy1956}.

\section{Thermal-Field emission models \label{sec:Model}}
We consider a generalized thermal-field electron emission model in the form of
\begin{equation}\label{eqn:J_perp_1}
    \mathcal{J}_\perp = \frac{eg}{ (2\pi)^3  } \int  v_\perp(\varepsilon_\mathbf{k}) \Theta\left( k_\perp \right)  f(\varepsilon_\mathbf{k})  \mathcal{T}(\varepsilon_\perp) \text{d}^3\mathbf{k} ,
\end{equation}
where $\mathcal{J}_{\perp}$ is the electron current density emitted vertically from a surface, $e$ is the charge of an electron, $g$ is the degeneracy factor, $f(\varepsilon_\mathbf{k})$ is the Fermi-Dirac distribution function and $\mathcal{T}(\varepsilon_\perp)$ is the tunnelling probability. 
The Heaviside step function, $\Theta\left( k_\perp \right)$, is defined to exclude the electronic states propagating backward in the negative vertical $\perp$ direction.
The group velocity along the emitting direction (denoted as $\perp$) is
\begin{equation}\label{eqn:grpvelocityGeneral}
    v_\perp(\varepsilon_\mathbf{k}) =\frac{1}{\hbar} \pdv{\varepsilon_\mathbf{k}}{\abs{k_\perp}} .
\end{equation}

\begin{table*}
\caption{Material parameters of the DSM: \ch{Cd3As2} and \ch{Na3Bi}, and WSMs: \ch{TaAs} and \ch{TaP}, are listed in this table. The parameters consists of the work function, $\Phi$, the intrinsic Fermi level, $\varepsilon_{F0}$, the Fermi velocities in the transverse direction, $v_{Fx}$ and $v_{Fy}$ and, the degeneracies of the nodes in the dispersion space, $g$. WSMs must have at least a pair of distinct Weyl nodes \cite{Wehling2014}. \label{tab:MaterialPara}}
\begin{ruledtabular}
\begin{tabular}{cccccc}
\textrm{DSMs/WSMs}&
$\Phi$ (eV)&
$\varepsilon_{F0}$ (eV)&
$v_{Fx}$ ($10^6$ m/s)&
$v_{Fy}$ ($10^6$ m/s)&
$g$\\ 
\hline
Cd3As2\footnote{References \cite{Liu2014,Jenkins2016,Crassee2018,Huang2020}} & $4.5$   & $0.1$             & $1.3$          & $1.28$          & $4$    \\
Na3Bi\footnote{References \cite{Liu2014a,Jenkins2016}}  & $2.35$  & $0.025$           & $0.374$        & $0.374$         & $4$    \\
TaAs\footnote{References \cite{Lee2015,Chi2017,Grassano2018}}   & $4.65$  & $-0.0221,-0.0089$ & $0.292,0.2055$ & $0.1532,0.1579$ & $8,16$ \\
TaP\footnote{References \cite{Lee2015,Grassano2018}}    & $5.405$ & $-0.0531,0.0196$  & $0.327,0.3736$ & $0.1594,0.2542$ & $8,16$ \\
\end{tabular}
\end{ruledtabular}
\end{table*}

In the following, we define $\perp$ to be pointing in the $z$ direction and $\parallel$ along the $x$-$y$ plane. 
The tunnelling probability $\mathcal{T}$ in \cref{eqn:J_perp_1} can be modelled after the SN barrier model \cite{Murphy1956,Forbes2006} to account for the image charge potential, given by
\begin{align}\label{eqn:SNTunnel}
    \mathcal{T}(\varepsilon_\perp) \approx \mathcal{D}_F \exp{\frac{\varepsilon_\perp - \varepsilon_F}{d_F}},
\end{align}
where $\mathcal{D}_F \equiv \exp\left( -b\mathtt{v}\sqrt{\Phi^3}/F \right)$ is the tunnelling exponent term and $d_F\equiv 2F/3b \sqrt{\Phi}\mathtt{t}$ is the decay width of the wave function through the barrier, $b = 4\sqrt{2m}/3e\hbar$ is a Fowler-Nordheim (FN) constant \cite{Forbes2007} with $m$ being the electron mass, $\Phi$ is the work function of the material, and $F$ is the applied field.
The image charge effect can be approximated with the following correction terms: \cite{Forbes2006}
\begin{align}\label{eqn:ImagecorrectionTerms}
    \mathtt{v}\approx 1-f_s + \frac{f_s}{6}\ln f_s, \quad \mathtt{t}\approx 1+\frac{f_s}{9}- \frac{f_s}{18}\ln f_s,
\end{align}
with $f_s = e^3F/4\pi\epsilon_0 \Phi^2$, where $\epsilon_0$ is the vacuum permittivity. 

\subsection{Generalized thermal-field emission current density for non-parabolic energy dispersion}

To accommodate the non-parabolic energy dispersion, we transform \cref{eqn:J_perp_1} into an alternative form. 
We first consider a generic energy dispersion, $\varepsilon_\mathbf{k}(\varepsilon_\parallel (\mathbf{k}_\parallel), \varepsilon_\perp(\mathbf{k}_\perp))$
with $\varepsilon_\parallel (\mathbf{k}_\parallel)$ and $\varepsilon_\perp(\mathbf{k}_\perp)$ as the energy component transverse and along the tunnelling direction, respectively. 
By rewriting \cref{eqn:J_perp_1} in terms of the $\perp$ and the $\parallel$ components through $\text{d}^3\mathbf{k} = k_\parallel \text{d}k_\parallel \text{d}\phi_{\mathbf{k}_\parallel} \text{d}k_\perp$, with $\phi_{\mathbf{k}_\parallel} = \tan^{-1}(k_y/k_x)$, we have
\begin{linenomath}
\begin{align}\label{eqn:J_perp}
    \mathcal{J}_\perp &=\frac{e}{2\pi\hbar} \left(\int f(\varepsilon_\mathbf{k})\mathcal{T}(\varepsilon_\perp) \Lambda_\perp \text{d}\varepsilon_\perp\right) \nonumber \\ &\times\left(\iint \frac{g}{(2\pi)^2} \Lambda_\parallel^{-1} \text{d}\phi_{\mathbf{k_\parallel}} \text{d}\varepsilon_\parallel \right),
\end{align}
\end{linenomath}
where we have defined the energy dispersion factors as
\begin{equation}\label{eqn:NP}
    \left(\Lambda_\perp, \Lambda_\parallel \right) \equiv \left(\pdv{\varepsilon_\mathbf{k}}{\abs{\varepsilon_\perp}}, \frac{1}{k_\parallel}\frac{\text{d} \varepsilon_\parallel}{\text{d} k_\parallel} \right),
\end{equation}
and the following transformation identity is used
$$\pdv{\varepsilon_\mathbf{k}}{\abs{k_\perp}}\mathbf{k}_\parallel \text{d}k_\parallel \text{d}k_\perp = \pdv{\varepsilon_\mathbf{k}}{\abs{\varepsilon_\perp}} d\varepsilon_\perp \frac{k_\parallel}{\partial\varepsilon_\parallel/\partial k_\parallel}d\varepsilon_\parallel.$$
Note that $k_\parallel = k_\parallel (\varepsilon_\parallel, \phi_{\mathbf{k}_\parallel})$ is defined. 
The energy dispersion factors in \cref{eqn:NP} play crucial roles in understanding the electron field emission with non-parabolic energy dispersion.

Consider an isotropic $3$D parabolic energy dispersion, $\varepsilon_\mathbf{k} = \hbar^2 (k_\parallel^2+k_\perp^2)/2m^*$ where $m^*$ is the electron effective mass and $\mathbf{k} = (\mathbf{k}_\parallel, \mathbf{k}_\perp)$ with $\mathbf{k}_\parallel$ representing the wave vector component transverse to the emission direction, the total energy is \emph{partitioned} into the emission component $\varepsilon_\perp \equiv \hbar^2 k_\perp^2/2m^*$, the transverse component $\varepsilon_\parallel \equiv \hbar^2 k_\parallel^2/2m^*$, and $\varepsilon_\mathbf{k} = \varepsilon_\parallel + \varepsilon_\perp$. 
In doing so, \cref{eqn:NP} becomes
\begin{equation}\label{eqn:parabolic}
    \left(\Lambda_\perp, \Lambda_\parallel \right)^{\text{parabolic}} \equiv \left(1, \frac{\hbar^2}{m^*}  \right),
\end{equation}
which is a constant term (independent of $\varepsilon_\parallel$ and $\varepsilon_\perp$).
Solving \cref{eqn:J_perp_1} with \cref{eqn:parabolic} will yield the classic Fowler-Nordheim (FN) law for cold field emission, and the Murphy-Good (MG) model for thermal-field emission. 

For a non-parabolic (or linear) energy dispersion, we have 
\begin{align}\label{eqn:GenEnergyDisp}
    \varepsilon_\mathbf{k} =
                \sqrt{\varepsilon_{\mathbf{k}_\perp}^2 + \varepsilon_{\mathbf{k}_\parallel}^2}.
\end{align}
For topological Dirac/Weyl semimetals, the quasiparticles around the Dirac point(s) are described by the effective Hamiltonian, $\mathcal{H}_\mathbf{k} = \hbar \sum_{i,a} v_{Fi,a} k_{i,a} \sigma_i$ where $i = \{x,y,z\}$ and $\sigma_i$ is the Pauli matrix along $\hat{i}$ and the subscript $a$ labels the $a$th Dirac cone \cite{Wehling2014,Lee2015,Chi2017,Du2020,Berry2020}. 
The energy dispersion of these topological semimetals (SM) is
\begin{align}\label{eqn:Multi-EkDisp}
    \varepsilon^{\textrm{SM}}_{\mathbf{k}} = \sum_{a=1}^N  \sqrt{\sum_i \hbar^2 v_{Fi,a}^2 k_{i,a}^2},
\end{align}
where $\varepsilon_{\parallel,a} \equiv \hbar  \left(v_{Fx,a}k_{x,a}^2 + v_{Fy,a}k_{y,a}^2\right)^{1/2}$ and $\varepsilon_{\perp,a} \equiv \hbar v_{Fz,a} k_{z,a}$. 
The corresponding energy dispersion factors for non-parabolic dispersion are
\begin{align}\label{eqn:linearSM}
    \left(\Lambda_{\perp,a}, \Lambda_{\parallel,a} \right)^{\text{SM}} \equiv \left(\frac{\abs{\varepsilon_{\perp,a}}}{\varepsilon_{\mathbf{k},a}}, \frac{\hbar^2 \tilde{v}^2_a}{\varepsilon_{\parallel,a}} \right),
\end{align}
where $\tilde{v}_a \equiv v_{Fy,a} \sqrt{(v_{Fx,a}/v_{Fy,a})^2\cos^2\phi_{\mathbf{k}_\parallel,a} + \sin^2\phi_{\mathbf{k}_\parallel},a}$ is the Fermi velocity along the $\parallel$ direction. This is in stark contrasts to the parabolic dispersion case as shown in \cref{eqn:parabolic} and will lead to a drastically different thermal-field emission characteristics for topological Dirac/Weyl semimetals to be reported below.

%
%

\begin{figure*}
    \begin{subfigure}[t]{1\textwidth}
        \includegraphics[width=1\textwidth]{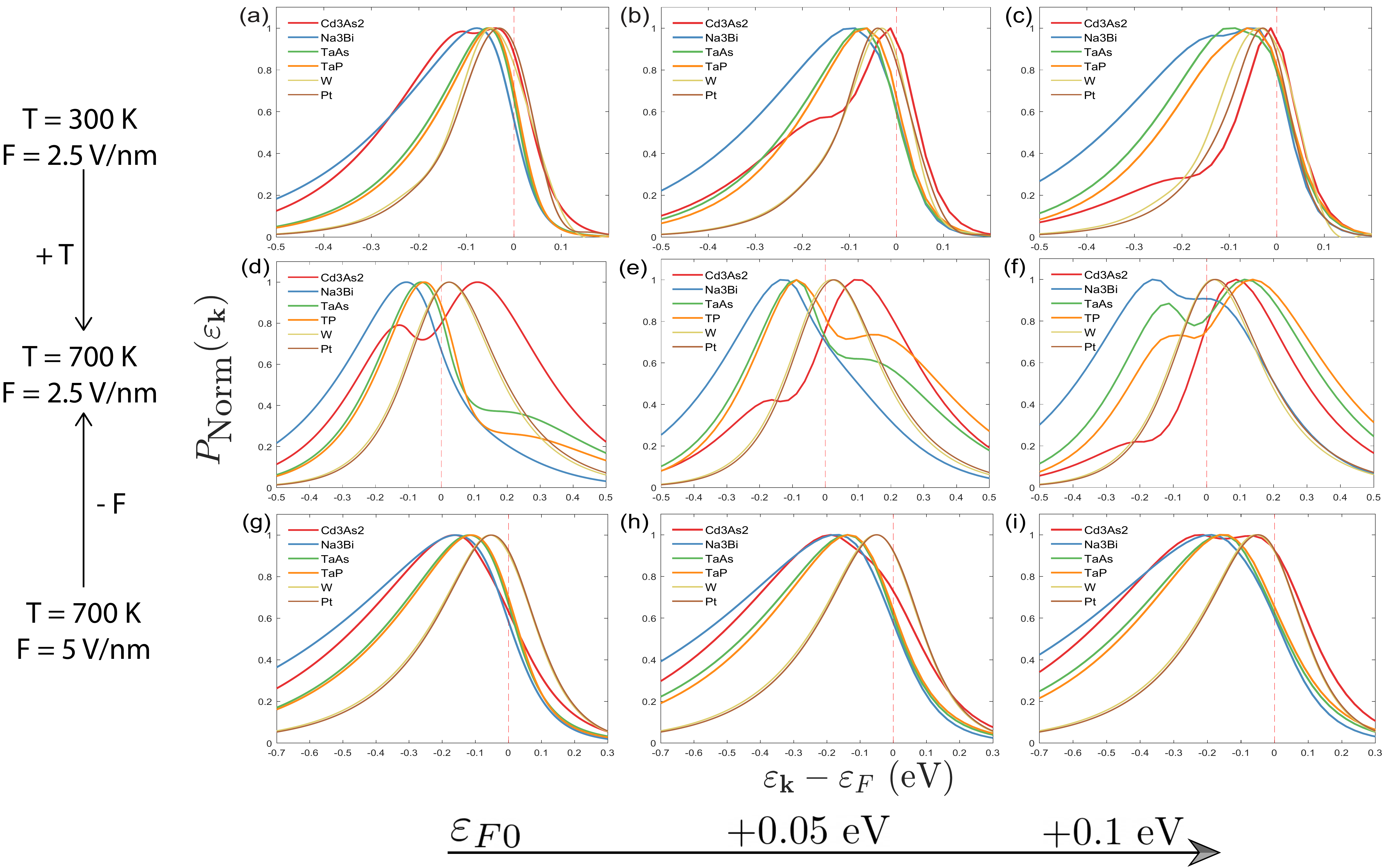}
        \phantomcaption
        \label{fig:2a}
    \end{subfigure}
    \begin{subfigure}[t]{0\textwidth}
        \includegraphics[width=0\textwidth]{Fig2.pdf}
        \phantomcaption
        \label{fig:2b}
    \end{subfigure}
    \begin{subfigure}[t]{0\textwidth}
        \includegraphics[width=0\textwidth]{Fig2.pdf}
        \phantomcaption
        \label{fig:2c}
    \end{subfigure}
    \begin{subfigure}[t]{0\textwidth}
        \includegraphics[width=0\textwidth]{Fig2.pdf}
        \phantomcaption
        \label{fig:2d}
    \end{subfigure}
    \begin{subfigure}[t]{0\textwidth}
        \includegraphics[width=0\textwidth]{Fig2.pdf}
        \phantomcaption
        \label{fig:2e}
    \end{subfigure}
    \begin{subfigure}[t]{0\textwidth}
        \includegraphics[width=0\textwidth]{Fig2.pdf}
        \phantomcaption
        \label{fig:2f}
    \end{subfigure}
    \begin{subfigure}[t]{0\textwidth}
        \includegraphics[width=0\textwidth]{Fig2.pdf}
        \phantomcaption
        \label{fig:2g}
    \end{subfigure}
    \begin{subfigure}[t]{0\textwidth}
        \includegraphics[width=0\textwidth]{Fig2.pdf}
        \phantomcaption
        \label{fig:2h}
    \end{subfigure}
    \begin{subfigure}[t]{0\textwidth}
        \includegraphics[width=0\textwidth]{Fig2.pdf}
        \phantomcaption
        \label{fig:2i}
    \end{subfigure}
\caption{The normalized (with respect to the maxima) TEDs are shown for \ch{Cd3As2} (red line), \ch{Na3Bi} (blue line), \ch{TaAs} (green line), \ch{TaP} (orange line), \ch{W} (yellow line) and \ch{Pt} (brown line) using material parameters in \cref{tab:MaterialPara} and \cite{Gadzuk1973,Nicolaou1975,Bordoloi1983} respectively. The red dotted line indicates the Fermi level. Along each row, the TEDs are plotted under $T = 300$ K and $F = 2.5$ V$/$nm for (a) to (c), $T = 700$ K and $F = 2.5$ V$/$nm for (d) to (f), $T = 700$ K and $F = 5$ V$/$nm for (g) to (i). Along each column, the TEDs are plotted using the intrinsic Fermi level $\varepsilon_{F0}$ in (a), (d), (g),  $\varepsilon_{F0} + 0.05$ eV in (b), (e), (h), and $\varepsilon_{F0}+0.1$ eV in (c), (f), (i). The TEDs are shifted along the $\varepsilon_\mathbf{k}$ axis such that all the Fermi level is zero (red vertical dotted line). The carriers left/right of the Fermi level are from either the valence/conduction band. The dual peak feature can be seen in all DSMs/WSMs, particularly in (d), (e) and (f) as we elevate $T$ and lower $F$ to sufficient values. Note the $2$ bulk materials (\ch{W} and \ch{Pt}) will always have single-peak TED}   
\label{fig:2}
\end{figure*}

\begin{figure*}
    \begin{subfigure}[t]{0.82\textwidth}
        \includegraphics[width=1\textwidth]{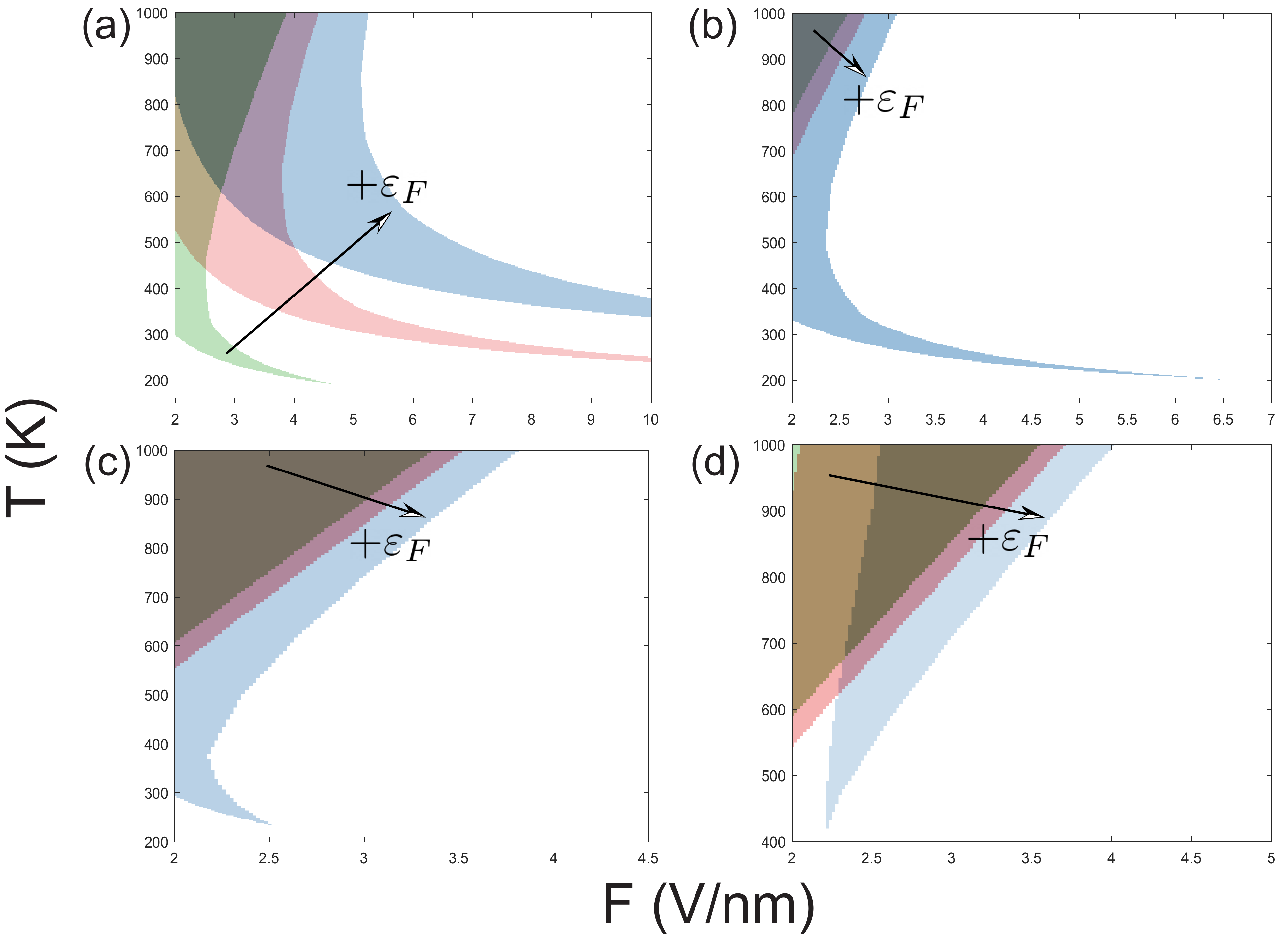}
        \phantomcaption
        \label{fig:3a}
    \end{subfigure}
    \begin{subfigure}[t]{0\textwidth}
        \includegraphics[width=0\textwidth]{Fig3.pdf}
        \phantomcaption
        \label{fig:3b}
    \end{subfigure}
    \begin{subfigure}[t]{0\textwidth}
        \includegraphics[width=0\textwidth]{Fig3.pdf}
        \phantomcaption
        \label{fig:3c}
    \end{subfigure}
    \begin{subfigure}[t]{0\textwidth}
        \includegraphics[width=0\textwidth]{Fig3.pdf}
        \phantomcaption
        \label{fig:3d}
    \end{subfigure}
\caption{The TED peak diagrams are shown for (a) \ch{Cd3As2}, (b) \ch{Na3Bi}, (c) \ch{TaAs}, and (d) \ch{TaP} based on the material parameters in \cref{tab:MaterialPara}. The coloured regions are the regions with dual-peak feature at different Fermi energy level from the intrinsic value $\varepsilon_{F0}$ in (green region) to $\varepsilon_{F0} + 0.05$ eV (red region) and $\varepsilon_{F0} +0.1$ eV (blue region). The white (uncoloured) regions are parameters that does not exhibit the dual peak feature.}   
\label{fig:3}
\end{figure*}

\subsection{Thermal-field emission current density and total energy distribution from Dirac cones\label{subsec:DiracCone}}

For an isotropic parabolic energy dispersion, \cref{eqn:J_perp} can be simplified as
\begin{align}\label{eqn:J_Para_1}
    \mathcal{J}_\perp^{\textrm{parabolic}} &=\frac{g_{s} m e}{(2\pi)^2\hbar^3} \int f(\varepsilon_\mathbf{k})\mathcal{T}(\varepsilon_\perp) \text{d}\varepsilon_\perp \text{d}\varepsilon_\mathbf{k},
\end{align}
$g_{s}$ is the spin degeneracy, $\pdv{\varepsilon_\mathbf{k}}{\varepsilon_\mu}=1$, $\mu \in \{\perp,\parallel\}$ which can be approximately solved to yield the well-known Murphy-Good (MG) thermal-field emission model \cite{Murphy1956}:
\begin{align}\label{eqn:J_MG}
    \mathcal{J}_\perp^{\textrm{MG}}&=\frac{a_{\textrm{FN}}F^2}{\Phi t^2}D_F \frac{k_B T \pi /d_F}{\sin{(k_B T \pi/d_F)}},
\end{align}
where $a_{FN} =e^3/(16\pi^2 \hbar) $ is the FN constant and $k_B$ is the Boltzmann constant.

The FN plot (for FN scaling) can be obtained by rearranging \cref{eqn:J_MG} such that, 
\begin{align}
    \ln\left(\frac{\mathcal{J}^{\textrm{MG}}_\perp}{F^2}\right) =-\frac{b\mathtt{v}\sqrt{\Phi^3}}{F}+ \ln\left(\frac{a_{\textrm{FN}} k_BT\pi/d_F}{\Phi \mathtt{t}^2\sin{(k_BT\pi/d_F)}}\right). \label{eqn:FNMG}
\end{align}
At $T=0$ K (cold field emission), it recovers the classical FN scaling of $\ln\left(\mathcal{J}_\perp^{\text{MG}}/F^2\right) \propto -1/F$.
Note the $F^2$ component in the logarithm term is a signature of field emission from bulk materials.

In contrast, \cref{eqn:J_perp} of a DSM/WSM exhibits a non-trivial difference due to \cref{eqn:linearSM}, which can be written as
\begin{align}\label{eqn:J_linear}
    \mathcal{J}_\perp^{\textrm{SM}} =\sum_{a=1}^N \frac{g_a e}{(2\pi\hbar)^3}& \left(\int\abs{\varepsilon_{\perp,a}} \mathcal{T}(\varepsilon_{\perp,a})f(\varepsilon_{\mathbf{k},a}) \text{d}\varepsilon_{\perp ,a}\right)\nonumber\\ &\times\left(\iint \frac{\varepsilon_{\parallel,a}}{\tilde{v}^2_a} \text{d}\phi_{\mathbf{k_\parallel},a} \text{d}\varepsilon_{\mathbf{k},a}\right),
\end{align}
where $g_a$ is the spin and node degeneracy for each contributing Dirac cone, $N$ is the number of contributing Dirac cone(s) to the emission.
\cref{eqn:J_linear} can be further simplified and numerically solved as
\begin{align}\label{eqn:J_Single}
    \mathcal{J}_\perp^{\textrm{SM}}&= \sum_{a=1}^N \frac{c_{\textrm{SM},a}F^2}{\Phi t^2}D_F\exp{-\frac{\varepsilon_F}{d_F}}\nonumber\\
    &\times\int_{-\infty}^{\infty} f(\varepsilon_{\mathbf{k},a})  \lambda(\varepsilon_{\mathbf{k},a}/d_F)\text{d}\varepsilon_{\mathbf{k},a},
\end{align}
where $c_{\textrm{SM},a} = e^3 g_a /(32m_e\pi^2\hbar v_{Fx,a}v_{Fy,a})$ is a constant with the rest mass of the electron being $m_e$ and
the dimensionless tunnelling function $\lambda$ is defined as
\begin{align}\label{eqn:lambda}
    \lambda(\varepsilon_{\mathbf{k},a}/d_F) &= \exp{\frac{\varepsilon_{\mathbf{k},a}}{d_F}}\left(\frac{\varepsilon_{\mathbf{k},a}}{d_F}-1\right)\text{sign}(\varepsilon_{\mathbf{k},a})\nonumber\\  &+ 1 + \text{sign}(\varepsilon_{\mathbf{k},a}).
\end{align}
The FN plot now takes the form of 
\begin{align}
    \ln\left(\frac{\mathcal{J}^{\textrm{SM}}_\perp}{F^2}\right) &= \sum_{a=1}^N  -\frac{b\mathtt{v}\sqrt{\Phi^3}}{F}- \frac{\varepsilon_F}{d_F}+\ln\left(\frac{c_{\textrm{SM},a}}{\Phi t^2}\right)\nonumber\\
    &+\ln\left(\int_{-\infty}^\infty f(\varepsilon_{\mathbf{k},a})  \lambda(\varepsilon_{\mathbf{k},a}/d_F)\text{d}\varepsilon_{\mathbf{k},a} \right).\label{eqn:FNSM}
\end{align}
At $T=0$ K (cold field emission), \cref{eqn:J_Single} becomes
\begin{align}
    \mathcal{J}_\perp^{\textrm{SM}}(T=0)&= \sum_{a=1}^N \frac{2c_{\textrm{SM},a}\sqrt{\Phi}D_F}{3b_{FN} t^2}\left(2\Gamma- + \frac{\varepsilon_F}{d_F}\Gamma_+\right)F^3,
\end{align}
where $\Gamma_\pm = 2\exp{-\varepsilon_F/d_F}\pm 1$. The scaling law at $T=0$ then becomes
\begin{align}
    \ln(\frac{\mathcal{J}_\perp^{\textrm{SM}}(T=0)}{F^3}) &= \sum_{a=1}^N -\frac{bv\sqrt{\Phi^3}}{F}+\ln\left(\frac{2c_{\textrm{SM},a}\sqrt{\Phi}}{ 3b_{FN}t^3}\right)\nonumber\\&+\ln\left( 2\Gamma_- + \frac{\varepsilon_F}{d_F}\Gamma_+ \right),
\end{align}
where $\Gamma_\pm$ is negligible in the pure field emission regime. 
Thus this produces an unexpected scaling of 
\begin{align}
   \ln\left(\frac{\mathcal{J}_\perp^{\text{SM}}}{F^3}\right) \propto -\frac{1}{F},\label{eqn:SMScale}
\end{align}
which is different from the classical FN scaling of $\ln\left(\mathcal{J}_\perp^{\text{MG}}/F^2\right) \propto -1/F$ for bulk solids with parabolic dispersion.

The energy spectrum of the emitted electrons is determined from the total energy distribution (TED): $P(\varepsilon_\mathbf{k})$ by considering $\partial\mathcal{J}_\perp / \partial\varepsilon_\mathbf{k}$ such that,
\begin{align}
    P(\varepsilon_\mathbf{k})= \frac{e}{2\pi\hbar^3} f(\varepsilon_\mathbf{k}) \int \mathcal{T}(\varepsilon_\perp) \Lambda_\perp D(\varepsilon_\mathbf{k}) \text{d}\varepsilon_\perp \text{d}\phi_{\mathbf{k}_\parallel}, \label{eqn:TEDGen}
\end{align}
where the transverse density of states is $\int D(\varepsilon_\parallel) \text{d}\phi_{\mathbf{k}_\parallel}\text{d}\varepsilon_\parallel = \int D(\varepsilon_\mathbf{k})\text{d}\phi_{\mathbf{k}_\parallel} \text{d}\varepsilon_\mathbf{k}$, with $D(\varepsilon_\mathbf{k}) = g/((2\pi)^2\Lambda_\parallel(\varepsilon_\mathbf{k}))$. 

For a parabolic dispersion, the TED takes on a conventional single peak behaviour due to the trivial energy dispersion factors in \cref{eqn:parabolic}, which gives us the well-known MG TED in the form of,
\begin{align}
    P^{\textrm{MG}}(\varepsilon_\mathbf{k})&=c_T \frac{F}{\sqrt{\Phi}t}D_F f(\varepsilon_\mathbf{k}) \exp{\frac{\varepsilon_\mathbf{k} - \varepsilon_F}{d_F}}.\label{eqn:TED_MG}
\end{align}

\begin{figure*}
    \begin{subfigure}[t]{1\textwidth}
        \includegraphics[width=\textwidth]{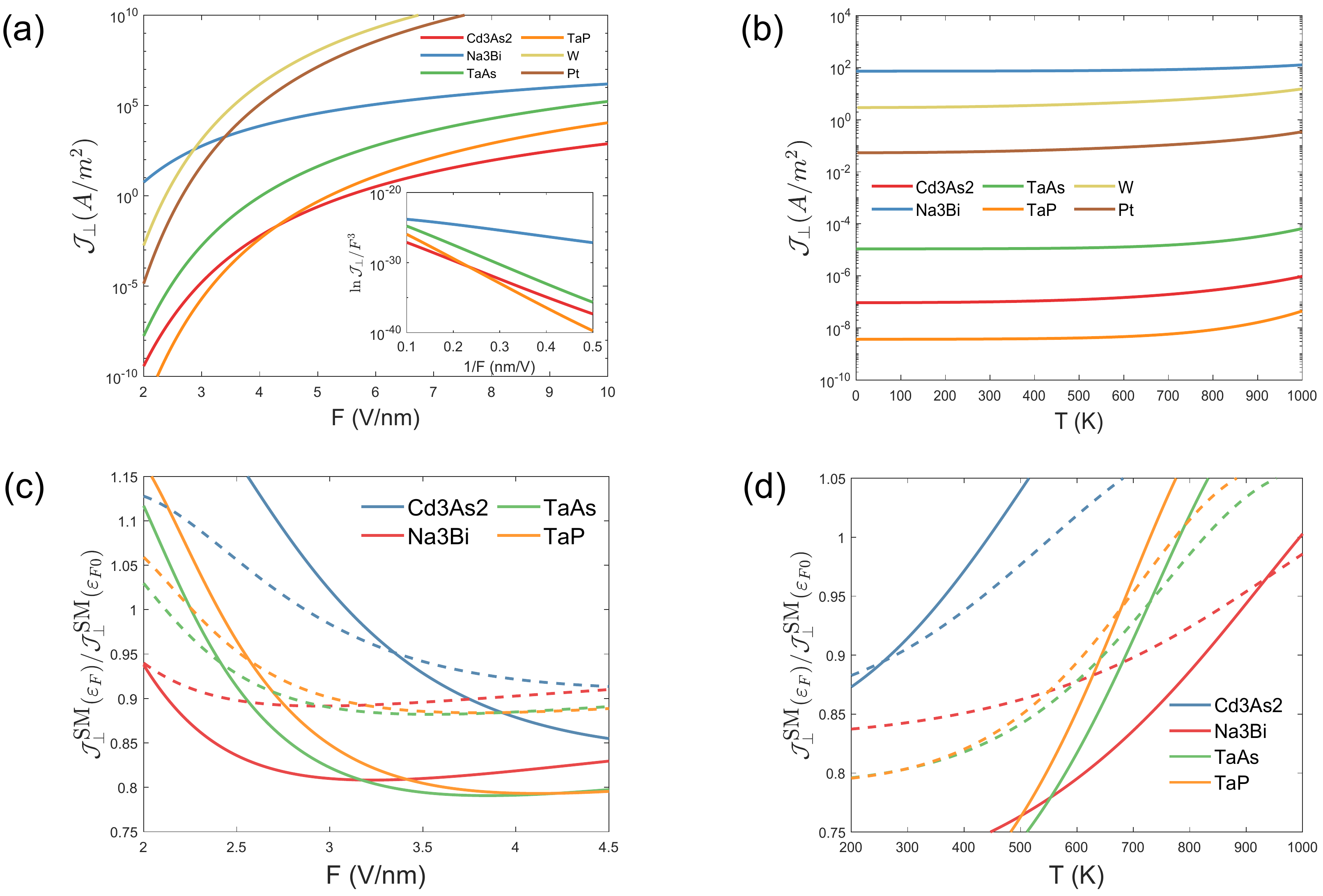}
        \phantomcaption
        \label{fig:4a}
    \end{subfigure}
    \begin{subfigure}[t]{0\textwidth}
        \includegraphics[width=\textwidth]{Fig4.pdf}
        \phantomcaption
        \label{fig:4b}
    \end{subfigure}
    \begin{subfigure}[t]{0\textwidth}
        \includegraphics[width=\textwidth]{Fig4.pdf}
        \phantomcaption
        \label{fig:4c}
    \end{subfigure}
    \begin{subfigure}[t]{0\textwidth}
        \includegraphics[width=\textwidth]{Fig4.pdf}
        \phantomcaption
        \label{fig:4d}
    \end{subfigure}
    \caption{Thermal-field emission current density at (a) $T$ = $700$ K with varying $F$, and (b) $F$ = $2.5$ V$/$nm with varying $T$, for $6$ materials: Tungsten (\ch{W}), Platinum (\ch{Pt}), \ch{Cd3As2}, \ch{Na3Bi}, \ch{TaAs} and \ch{TaP}. The inset in (a) shows the unconventional $F^3$ scaling law in the FN plots for DSMs/WSMs. (b) The ratio of the emission current density at enhanced Fermi energy level ($\varepsilon_F$) to that at intrinsic level ($\varepsilon_{F0}$) at (c) $T$ = $700$ K with varying $F$ and (d) at $F$ = $2.5$ V$/$nm with varying $T$. The dashed and solid lines corresponds to the enhanced $\varepsilon_F$ of $0.05$ eV and $0.1$ eV, respectively.}
    \label{fig:4}
\end{figure*}

On the contrary, \cref{eqn:linearSM} will produce a differing and non-trivial ($\varepsilon_\parallel$ and $\varepsilon_\perp$ dependence) term in the transverse DOS and the $\varepsilon_\perp$ integration, which is
\begin{align}
    P^{\textrm{SM}}(\varepsilon_\mathbf{k})&=\sum_{a=1}^N \frac{c_{\textrm{SM},a}F^2}{\Phi t^2}D_F \exp{-\frac{\varepsilon_F}{d_F}}\nonumber\\ &\times f(\varepsilon_{\mathbf{k},a})  \lambda(\varepsilon_{\mathbf{k},a}/d_F)\label{eqn:TED_SM}.
\end{align}

Unlike the parabolic dispersion commonly used for bulk metals in \cref{fig:1b}, the linear dispersion and the vanishing density of states of a DSM/WSM portrays a new dual-peak feature in the TED, as illustrated in \cref{fig:1c}.
This distinctive feature is attributed by the products of three terms in \cref{eqn:TEDGen}: $f(\varepsilon_\mathbf{k})$ (green solid line), the $\varepsilon_\perp$ integration of $\Lambda_\perp \mathcal{T}(\varepsilon_\perp)$ (yellow dotted line), and the $\phi_{\mathbf{k}_\parallel}$ integration of $D(\varepsilon_\parallel)$ (red dotted line).
The coloured region under the TED graph corresponds to the emitting energy band, where blue/orange represents the emission from the valence/conduction band. 
The purple region represents the emission around the Fermi level. 
As seen below, this unconventional behaviour in the TED can be utilised to manipulate the energy profile of the emission process.

\section{Results and Discussion \label{sec:R&D}}
\subsection{Dual-peak total energy distribution}

We investigate the appearance of the dual-peak feature in $4$ topological semimetals in \cref{fig:2}.
The material parameters for the topological semimetals are given in \cref{tab:MaterialPara}. 
In comparison, $2$ bulk metals \cite{Gadzuk1973,Nicolaou1975,Bordoloi1983} are considered in our calculations: (a) Tungsten \ch{W} with $\Phi = 5.25$ eV and $\varepsilon_{F0} = 11.55$ eV;  (b) Platinum \ch{Pt} with $\Phi = 5.65$ eV and $\varepsilon_{F0} = 8.78$ eV.
Unlike topological semimetals, materials modelled with a parabolic dispersion like \ch{W} and \ch{Pt} do not generate this duel-peak behaviour in its TED for all ranges of $F$, $T$ and $\varepsilon_F$.
This behaviour arises in topological semimetals from $P^{\textrm{SM}}(\varepsilon_\mathbf{k}) \propto \lambda(\varepsilon_\mathbf{k}/d_F)$ in \cref{eqn:TED_SM} due to the differing (Dirac conic) band structure embedded in \cref{eqn:linearSM}.
In contrast, this is not seen for a parabolic dispersion as $P^{\textrm{MG}}(\varepsilon_\mathbf{k}) \propto \exp{\varepsilon_\mathbf{k}/d_F}$ in \cref{eqn:TED_MG} from \cref{eqn:parabolic} and is consistent with \cref{fig:1b,fig:1c}.  
The dual-peak TED is thus a signature of the Dirac conic band structure.

Furthermore, the dual-peak feature can be generated or removed by regulating the energies of the carriers by changing $T$ and $F$.
This can be seen by comparing \crefrange{fig:2a}{fig:2c} ($T=300$ K ,$F=2.5$ V$/$nm) and \crefrange{fig:2g}{fig:2i} ($T=700$ K, $F=5$ V$/$nm) against \crefrange{fig:2d}{fig:2f} ($T=700$ K, $F=2.5$ V$/$nm).
The dependency of $f(\varepsilon_\mathbf{k})$ in \cref{eqn:TED_SM} regulates the number of high energy carriers (from the conduction band) as $T$ is varied.
As the height of the potential barrier is varied through $F$, the dependency of $D_F \lambda(\varepsilon_\mathbf{k}/d_F)$ in \cref{eqn:TED_SM} makes it easier/harder for the low energy carriers (from the valence band) to tunnel across the surface barrier.
Hence, the dual-peak feature can be modulated by tuning the $F$ and $T$ until the number of high and low energy carriers for emission is similar/vastly different. 
This feature appears in \cref{fig:2d} for \ch{Cd3As2} (red line), with the peak before/after the Fermi level (red dotted line) termed as the low/high energy peak.
Likewise, the dual peaks disappear as $F$ is further increased, allowing more low energy carriers to be emitted as shown in \ch{Cd3As2} [\cref{fig:2f}].
Thus, the dual-peak feature is not robust against the variation of $T$ and $F$.

In addition to $F$ and $T$, the dual-peaks can also be created by varying $\varepsilon_F$. 
This can be seen across the columns of \cref{fig:2}, that an increase of $\varepsilon_F$ introduces more hot carriers.
This is supported by \cref{eqn:TED_SM}, where the $f(\varepsilon_\mathbf{k})$ term introduce more hot carriers and the $\exp{-\varepsilon_F/d_F} \lambda(\varepsilon_\mathbf{k}/d_F)$ term allows the hot carriers to tunnel through easier from increasing $\varepsilon_F$.
An explicit example can be seen in \cref{fig:2e,fig:2f} for \ch{Na3Bi} (blue line), where the dual-peak feature appears only after a sufficient increased of $\varepsilon_F$.
Hence, the $\varepsilon_F$ plays a vital role in regulating the number of hot carriers for emission, which is crucial in manipulating the dual-peak feature. 

\subsection{Susceptibility diagram of the dual-peak TED}
In \cref{fig:3}, we further study the mechanism of the dual-peak feature in \cref{fig:2} as illustrated by increasing $\varepsilon_F$ in the susceptibility ($F$-$T$) diagram of the dual-peak feature.
It can be seen that the dual-peak regions expand and shift with increasing $\varepsilon_F$ in all four topological semimetals. 
This is predicted in \cref{eqn:TED_SM,fig:2}, where the increase in $\varepsilon_F$ increases the number of high energy carriers.
Due to the scarcity of the high energy carriers in the field emission regime, the expansion of the region is expected as the increased number of high energy carriers makes it more conducive to generate the dual-peak feature by the variation of $F$ and $T$, which is seen in \cref{fig:3b,fig:3c} for \ch{Na3Bi} and \ch{TaAs}. 
The shifting of the regions indicates that the dual-peak feature can be generated or removed at different ranges of $F$ and $T$, with an increasing number of high energy carriers by increasing $\varepsilon_F$.
This is explicitly seen in \cref{fig:3a,fig:3d} for \ch{Cd3As2} and \ch{TaP} and is supported by the gain or lost of the dual-peak feature in \cref{fig:2}.
Thus, the susceptibility diagram pinpoints the parameters to achieve the dual-peak feature, which can be useful in generating a larger emission current density as seen in \cref{fig:4}.

Furthermore, the susceptibility diagram shows that the dual-peak feature can be easily observed for \ch{Cd3As2} with its intrinsic Fermi level, $\varepsilon_{F0}$.
Not only does \ch{Cd3As2} have the largest green ($\varepsilon_{F0}$) region, it also have a very noticeable shift in \cref{fig:3a}, which is further supported by \cref{fig:2}, where it can exhibits this feature more prominently.
This is due to \ch{Cd3As2} having high $\varepsilon_{F0}$ as compared to the other three topological semimetals.
Thus, the susceptibility diagram shows that \ch{Cd3As2} is a suitable candidate amongst these four topological semimetals to observe this dual-peak feature in future experiments.

\subsection{Emission current density and scaling law}
For comparison, the thermal-field emission current density between DSM/WSM [$\mathcal{J}_\perp$ in \cref{eqn:J_Single}] and traditional materials with parabolic dispersion [\cref{eqn:J_MG}] is plotted in
\cref{fig:4a} at $T = 700$ K and \cref{fig:4b}  $2.5$ V$/$nm at its intrinsic Fermi level, $\varepsilon_{F0}$.
Due to the semi-metallic nature of DSMs/WSMs, it is expected that they exhibit a lower emission current density than the metallic emitters due to their limited electronic carrier density. 
Indeed, we observe that the emission current density ($\mathcal{J}_\perp$) of DSMs/WSMs [\ch{Cd3As2} (red line), \ch{TaAs} (green line) and \ch{TaP} (orange line)] is lower than traditional materials [\ch{W} (yellow line) and \ch{Pt} (brown line)] with the exception of \ch{Na3Bi} (blue line) due to its significant lower work function of $2.35$ eV.
Such behaviour is expected from \cref{eqn:J_MG,eqn:J_Single} as they share similar $F$ and $T$ dependence terms in their thermal-field emission current densities. 

Interestingly, the inset in \cref{fig:4a} shows that the field emission follows an unconventional scaling of $\ln(J/F^3) \propto -1/F$ (\cref{eqn:SMScale}) for all topological semimetals.
The excellent linearity of $\ln(J/F^3) \propto -1/F$ reveals that DSM/WSM emitter follows an unconventional non-FN scaling not commonly seen in traditional materials.

The effects of the parameters $\varepsilon_F$ on the ratio of $\mathcal{J}^{\text{SM}}_\perp(\varepsilon_F)$ against $\mathcal{J}^{\text{SM}}_\perp(\varepsilon_{F0})$ while varying $F$ and $T$ are being investigated in \cref{fig:4c,fig:4d}. 
The ratio dips below $1$ for increasing $F$ / decreasing $T$. 
With a lower $\varepsilon_F$, it is easier to activate more low energy carriers for emission as compared to a higher $\varepsilon_F$ with an enlarged conduction band when increasing $F$.
Similarly, the reduced number of high energy carriers has a lesser effect on the TED as with a lower $\varepsilon_F$, the emission is dominated by low energy carriers.
This goes in tandem with the loss of the dual-peak feature, which signifies the dominance of the low energy carriers for $\varepsilon_{F0}$ as seen either fixing $T=700$ K or $2.5$ V$/$nm in \cref{fig:3}.
Notably, \ch{Na3Bi} exceed $1$ in \cref{fig:4c}, which is expected as it did not exhibit the dual-peak property with $\varepsilon_{F0}$ as seen in \cref{fig:2,fig:3}. 
Thus, a higher thermal-field emission current densities can be achieved by either altering $\varepsilon_F$ or by manipulating the dual-peak feature with the variation of $F$ and $T$.

\section{Conclusion \label{sec:Con}}
In conclusion, we have developed a thermal-field emission model for $3$D Dirac semimetal (DSM) and Weyl semimetal (WSM) with a linear Dirac conic energy dispersion such as \ch{Cd3As2}, \ch{Na3Bi}, \ch{TaAs}, and \ch{TaP}.
Our results predict the existence of a non-trivial dual-peak feature in the total energy distribution (TED) and a new $\ln(J/F^3) \propto -1/F$ scaling law, which is absent in an electron field emitter composed of traditional metals with conventional $3$D parabolic energy band structure.
The characteristics of this dual-peak feature are unique to DSM/WSM and may serve as a smoking gun signature for the Dirac conic energy dispersion in topological semimetals.
Furthermore, the relatively low work function of \ch{Na3Bi} can be beneficial for field emission application due to its high emission current density.
\ch{Cd3As2} with its sufficiently high Fermi level (in spite of high work function), is also a suitable candidate for achieving a larger emission current density by exploiting the high sensitivity of the dual-peak feature with the variation of $F$, $T$ and $\varepsilon_F$.
Finally, we remark that our model does not capture secondary effects, such as field-induced topological phase transition \cite{Wehling2014,Wang2017,Arm2018}, band bending \cite{Lito2007}, space charge \cite{Khalid2016,Zhang2021}, and Fermi velocity shifting \cite{Grassano2018,Baba2019}.
Such effects could be included in future works to investigate their roles on the thermal-field emission characteristics of $3$D DSM/WSM.

\section{Acknowledgements}
This work is funded by MOE Tier 2 (2018-T2-1-007). Y.S.A. acknowledges the support of SUTD Startup Research Grant (Project No. SRT3CI21163). W.J.C. acknowledge MOE PhD RSS.

\appendix

\section{Derivation of TED and emission current density with linear (non-parabolic) dispersion for a DSM/WSM \label{Appendix:DeriveTEDJ_NPSM}}

The generalised emission electrical current density from a $3$D bulk electron emitter with a linear (non-parabolic) dispersion is shown in \cref{eqn:J_perp} of \cref{sec:Model} as 
\begin{linenomath}
\begin{align}
    \mathcal{J}^L_\perp &=\frac{e}{2\pi\hbar} \left(\int f(\varepsilon_\mathbf{k})\mathcal{T}(\varepsilon_\perp) \Lambda_\perp \text{d}\varepsilon_\perp\right) \nonumber \\ &\times\left(\iint \frac{g}{(2\pi)^2} \Lambda_\parallel^{-1} \text{d}\phi_{\mathbf{k_\parallel}} \text{d}\varepsilon_\parallel \right),
\end{align}
\end{linenomath}
where the superscript $L$ indicates a linear dispersion. Inserting \cref{eqn:linearSM} into \cref{eqn:J_perp} garner us \cref{eqn:J_linear} such that,
\begin{align}
    \mathcal{J}_\perp^{\textrm{SM}} =\sum_{a=1}^N \frac{g_a e}{(2\pi\hbar)^3}& \left(\int\abs{\varepsilon_{\perp,a}} \mathcal{T}(\varepsilon_{\perp,a})f(\varepsilon_{\mathbf{k},a}) \text{d}\varepsilon_{\perp ,a}\right)\nonumber\\ &\times\left(\iint \frac{\varepsilon_{\parallel,a}}{\tilde{v}^2_a} \text{d}\phi_{\mathbf{k_\parallel},a} \text{d}\varepsilon_{\mathbf{k},a}\right).
\end{align}

Without loss of generality, we can discard the Dirac node label, $a$ by specifying a Dirac node. This allow us to temporary remove the summation, which can be placed back at the end.

The $\varepsilon_{\perp}$ integration then gives:
\begin{align}\label{eqn:B2}
    \int\abs{\varepsilon_{\perp}}\mathcal{T}(\varepsilon_{\perp})\text{d}\varepsilon_{\perp} = \frac{D_F}{\hbar\varepsilon_{\mathbf{k}}} \int_{-\infty}^{\varepsilon_{\mathbf{k}}} \abs{\varepsilon_{\perp}} e^{\left(\varepsilon_{\perp} -\varepsilon_F\right) / d_F} \text{d}\varepsilon_{\perp},
\end{align}
where the approximation $(0 \rightarrow -\infty)$ can be made as the emission takes place near the Fermi level. 

Due to the absolute function in $\varepsilon_\perp$, we have to consider both the $\varepsilon_\mathbf{k} < 0$ and $\varepsilon_\mathbf{k} > 0$ regime, such that for $\varepsilon_\mathbf{k} < 0$,
\begin{linenomath}
\begin{align}\label{eqn:B3}
    &\int_{-\infty}^{\varepsilon_\mathbf{k}} \abs{\varepsilon_\perp} e^{\left(\varepsilon_\perp -\varepsilon_F\right) / d_F} \text{d}\varepsilon_\perp= d_F^2\exp{\frac{\varepsilon_\mathbf{k}-\varepsilon_F}{d_F}}\left(1-\frac{\varepsilon_\mathbf{k}}{d_F}\right),
\end{align}
\end{linenomath}
and for $\varepsilon_\mathbf{k} > 0$,
\begin{align}\label{eqn:B4}
    \int_{-\infty}^{\varepsilon_\mathbf{k}} \abs{\varepsilon_\perp} e^{\left(\varepsilon_\perp -\varepsilon_F\right) / d_F} \text{d}\varepsilon_\perp&=d_F^2\exp{-\frac{\varepsilon_F}{d_F}}\nonumber\\
    &\times\left(2+\left(\frac{\varepsilon_\mathbf{k}}{d_F}-1\right)\exp{\frac{\varepsilon_\mathbf{k}}{d_F}} \right).
\end{align}
Hence, we can combine them into a dimensionless tunnelling function such that
\begin{linenomath}
\begin{align}
    \lambda(\varepsilon_\mathbf{k}/d_F) = \exp{\frac{\varepsilon_\mathbf{k}}{d_F}}\left(\frac{\varepsilon_\mathbf{k}}{d_F}-1\right)\text{sign}(\varepsilon_\mathbf{k}) + 1 + \text{sign}(\varepsilon_\mathbf{k})
\end{align}
\end{linenomath}
where
$$
\lambda(\varepsilon_\mathbf{k}/d_F) = 
\begin{cases}
    \left(1-\frac{\varepsilon_\mathbf{k}}{d_F}\right)\exp{\frac{\varepsilon_\mathbf{k}}{d_F}}&,\varepsilon_\mathbf{k}<0 \\
    2+\left(\frac{\varepsilon_\mathbf{k}}{d_F}-1\right)\exp{\frac{\varepsilon_\mathbf{k}}{d_F}} &,\varepsilon_\mathbf{k} \geq 0.
\end{cases}
$$
This can be done as the limits of \cref{eqn:B3,eqn:B4} at $0$ are equal and can be easily verified.
Hence, the $\varepsilon_\perp$ integration becomes
\begin{align}
    \int\abs{\varepsilon_{\perp}}\mathcal{T}(\varepsilon_{\perp})\text{d}\varepsilon_{\perp} = \frac{D_F}{\hbar \varepsilon_\mathbf{k}} \lambda(\varepsilon_\mathbf{k}/d_F).\label{eqn:AppxPerpTerm}
\end{align}

Next for the $\varepsilon_\parallel$ and $\phi_{\mathbf{k}_\parallel,a}$ integration,
\begin{align}
    \iint \frac{\varepsilon_{\parallel}}{\tilde{v}^2} \text{d}\phi_{\mathbf{k_\parallel}} \text{d}\varepsilon_{\parallel}&= \int\frac{2\pi\varepsilon_\mathbf{k}}{ v_{Fx}v_{Fy}}\text{d}\varepsilon_\mathbf{k}\label{eqn:AppxParaTerm}
\end{align}
where $\tilde{v} \equiv v_{Fy} \sqrt{(v_{Fx}/v_{Fy})^2\cos^2\phi_{\mathbf{k}_\parallel} + \sin^2\phi_{\mathbf{k}_\parallel}}$ and the identity, $\int_0^{2\pi} (a^2\cos^2 \phi_{\mathbf{k}_\parallel} + \sin^2 \phi_{ \mathbf{k}_\parallel} )^{-1}d\phi_{\mathbf{k}_\parallel}  = 2\pi/a$ and $\partial \varepsilon_\mathbf{k} / \partial\varepsilon_\parallel = \varepsilon_\parallel/\varepsilon_\mathbf{k}$ was used.

This changes \cref{eqn:J_linear} into \cref{eqn:J_Single}
\begin{linenomath}
\begin{align}
    \mathcal{J}_\perp^{\textrm{SM}}&= \sum_{a=1}^N \frac{c_{\textrm{SM},a}F^2}{\Phi t^2}D_F\exp{-\frac{\varepsilon_F}{d_F}}\nonumber\\
    &\times\int_{-\infty}^{\infty} f(\varepsilon_{\mathbf{k},a})  \lambda(\varepsilon_{\mathbf{k},a}/d_F)\text{d}\varepsilon_{\mathbf{k},a}.
\end{align}
\end{linenomath}
where the $\textrm{SM}$ represents the emission current density of a topological semimetal, $c_{\textrm{SM}} = e^3 g/32m\pi^2\hbar v_{Fx}v_{Fy}$ is a constant, with $m_e$ being the rest mass of the electron. 

The TED can be extracted from the emission current density from 
\begin{align}
     P(\varepsilon_\mathbf{k})=\pdv{\mathcal{J}_\perp}{\varepsilon_\mathbf{k}}, 
\end{align}
which gives us a general form:
\begin{align}
    P(\varepsilon_\mathbf{k})=\frac{e}{2\pi\hbar^3} f(\varepsilon_\mathbf{k}) \int \mathcal{T}(\varepsilon_\perp) \Lambda_\perp D(\varepsilon_\mathbf{k}) \text{d}\varepsilon_\perp \text{d}\phi_{\mathbf{k}_\parallel}.
\end{align}
For DSMs/WSMs, we can apply the results from the various integration in \cref{eqn:AppxPerpTerm,eqn:AppxParaTerm} and \cref{eqn:linearSM} to obtain the TED in \cref{eqn:TED_SM} such that,
\begin{linenomath}
\begin{align}
    P^{\textrm{SM}}(\varepsilon_\mathbf{k})&=\sum_{a=1}^N \frac{c_{\textrm{SM},a}F^2}{\Phi t^2}D_F \exp{-\frac{\varepsilon_F}{d_F}}\nonumber\\ &\times f(\varepsilon_{\mathbf{k},a})  \lambda(\varepsilon_{\mathbf{k},a}/d_F).\label{eqn:AppxTED}
\end{align}
\end{linenomath}

\end{document}